\documentclass[12pt,english]{article}
\usepackage[T1]{fontenc}
\usepackage[latin9]{inputenc}
\usepackage{geometry}
\usepackage{array}
\usepackage{float}
\usepackage{textcomp}
\usepackage{multirow}
\usepackage{amsmath}
\usepackage{graphicx}
\usepackage[authoryear]{natbib}

\makeatletter

\providecommand{\tabularnewline}{\\}
\floatstyle{ruled}
\newfloat{algorithm}{tbp}{loa}
\providecommand{\algorithmname}{Algorithm}
\floatname{algorithm}{\protect\algorithmname}
\usepackage{tikz}
\usetikzlibrary{calc}

\@ifundefined{date}{}{\date{}}
\usepackage{graphicx, graphics, color, amsfonts, float}
\usepackage{hyperref}
\usepackage[T1]{fontenc}
\usepackage{natbib}
\usepackage{rotating}
\usepackage{amsmath}
\hypersetup{colorlinks=true}
\hypersetup{linkcolor=blue}
\hypersetup{urlcolor=blue}
\hypersetup{citecolor=blue}
\usepackage{lscape}
\usepackage{longtable}
\usepackage{booktabs,fixltx2e}
\usepackage{threeparttable}
\floatstyle{plaintop}
\restylefloat{table}
\usepackage{subfig}
\usepackage{xfrac}
\usepackage{cancel}
\usepackage[font=footnotesize, tableposition=bottom]{caption}
\usepackage{setspace}
\usepackage{algorithm}
\usepackage{pdflscape}
\makeatother

\usepackage{babel}
\begin{document}
\title{Variational inference for steady-state BVARs}
\author{Oskar Gustafsson and Mattias Villani}
\maketitle
\begin{center}
\emph{Department of Statistics, Stockholm University}
\par\end{center}
\begin{abstract}
The steady-state Bayesian vector autoregression (BVAR) makes it possible
to incorporate prior information about the long-run mean of the process.
This has been shown in many studies to substantially improve forecasting
performance, and the model is routinely used for forecasting and macroeconomic
policy analysis at central banks and other financial institutions.
Steady-steady BVARs are estimated using Gibbs sampling which is time-consuming
for the increasingly popular large-scale BVAR models with many variables.
We propose a fast variational inference (VI) algorithm for approximating
the parameter posterior and predictive distribution of the steady-state
BVAR, as well as log predictive scores for model comparison. We use
simulated and real US macroeconomic data to show that VI produces
results that are very close to results from Gibbs sampling. The computing
time of VI can be orders of magnitudes lower than Gibbs sampling,
in particular for log predictive scores, and VI is shown to scale
much better with the number of time series in the system.
\end{abstract}

\section{Introduction}

There is a clear trend in time series econometrics and forecasting
to include increasingly many predictors in combination with Bayesian
shrinkage priors. This is a sensible way to incorporate as much of
the information as possible into the analysis while still controlling
overparameterization to improve forecasting performance; see e.g.
\citet{banbura}, \citet{giannone2015}, and many others. Easily accessible
large-scale data and computing power has generated similar trends
in many other applied fields and is currently driving much of the
methodological developments in statistics and machine learning.

The computational burden of inference and prediction in large-scale
models has led researchers to make overly simplifying model or prior
assumptions to reduce computing times. For example, cross-lag shrinkage
in Bayesian vector autoregressions (BVARs) is an intuitive and important
hyperparameter for practitioners \citep{gustafsson2020bayesian},
but is often dropped as an option for computational reasons \citep{koop2013}.
Another example where it is hard to reduce the computing times by
model reduction is the steady-state BVAR of \citet{villani2009},
which uses a reformulation of the VAR with an informative prior on
the long-run mean level of the process. This model has proven very
useful for forecasting of macroeconomic variables, see e.g. \citet{beechey2010forecasting},
\citet{Gustafsson2016}, and \citet{stockhammar2017impact} and is
routinely used in many central banks and other finanicial institutions.
The Gibbs sampler in \citet{villani2009} is easy to implement but
becomes a bottleneck for large-scale models, in particular when performing
model comparison using predictive measures that requires running the
Gibbs sampling algorithm on many different subsets of the data.

\emph{Variational Inference} (VI), also called \emph{variational }Bayes,
is an optimization method for approximating probability densities
that originates in the machine learning literature, see  \citet{ormerod2010explaining}
and \citet{blei2017variational} for introductions for statisticians.
The method is widely used to approximate posterior densities in Bayesian
models, and can be seen as an alternative to Gibbs sampling, or more
generally, Markov chain Monte Carlo (MCMC) algorithms. VI has recently
been used to estimate large scale BVARs in e.g. \citet{koop2018variational}
and \citet{gefang2019variational}. There are benefits and drawbacks
with both MCMC/Gibbs sampling and VI. Posterior draws using MCMC are
known to converge in distribution to the target posterior as the number
of samples grows, but is typically computationally slow, especially
for models with many parameters. VI is instead an approximate method
based on optimization, with the advantage that it is typically much
faster than MCMC, and scales better to large data sets \citep{blei2017variational}.
Moreover, VI is especially fast when the model needs to be repeatedly
re-estimated on slightly extended datasets, as typically done when
evaluating forecasting performance and model comparison via log predictive
scores (LPS). Each VI optimization can then be initialized with very
good parameter values from the previous estimation and converges extremely
quickly, while MCMC methods always need to sample until convergence
\citep{nott2012regression}.

We develop a fast so-called structured mean field VI algorithm for
the steady-state BVAR model. The VI updates are shown to be available
in closed form, which makes the algorithm extremely robust and fast.
The algorithm is demonstrated on real and simulated data to be substantially
faster than the currently used Gibbs sampling in \citet{villani2009}
for approximating a single posterior, and orders of magnitudes faster
for model comparision using the LPS. The computing time for VI is
also demonstrated to scale much better with respect to the number
of time series compared to Gibbs sampling. Importantly, the VI approximation
is shown to be accurate for the typical applications of steady-state
BVAR used in practical work.

\section{BVARs and Variational Inference}

\subsection{Steady-state BVARs\label{subsec:Steady-state-BVARs}}

The steady-state BVAR model \citep{villani2009} is given by:

\begin{equation}
\begin{aligned}\mathbf{\Pi}(L)(\mathbf{y}_{t}-\Psi\mathbf{x}_{t}) & =\mathbf{\varepsilon}_{t},\qquad\text{where }\mathbf{\varepsilon}_{t}\overset{\text{iid}}{\sim}N(\mathbf{0},\Sigma),\end{aligned}
\label{SSBVAR}
\end{equation}
where $E[\mathbf{y}_{t}]=\Psi\mathbf{x}_{t}$ is the unconditional
mean of the process. We will for simplicity assume that $\mathbf{x}_{t}=1\enskip\forall t,$
but the presented method applies to any exogenous $\mathbf{x}_{t}$
vector, for example with dummy variables for level shifts. The extension
to the case of a latent mean process is discussed in Section \ref{sec:Discussion}.
Following \citet{villani2009}, we assume prior independence between
the parameter blocks and
\begin{equation}
\begin{aligned}p(\Sigma)\sim & |\Sigma|^{-(n+1)/2}\\
vec(\Pi)\sim & N(\underline{\boldsymbol{\theta}}_{\Pi},\underline{\Omega}_{\Pi})\\
\Psi\sim & N(\underline{\boldsymbol{\theta}}_{\Psi},\underline{\Omega}_{\Psi}),
\end{aligned}
\end{equation}
where $\underline{\boldsymbol{\theta}}_{\Psi}\text{ and }\underline{\Omega}_{\Psi}$
are the prior mean and covariance matrix for the steady-states. The
vector $\underline{\boldsymbol{\theta}}_{\Pi}$ is the mean of the
dynamic coefficients and the covariance matrix for the VARdynamics
$\underline{\Omega}_{\Pi}$ is diagonal with elements
\begin{equation}
\underline{\omega}_{ii}=\begin{cases}
\frac{\lambda_{1}^{2}}{(l^{\lambda_{3}})^{2}},\text{ for own lag }l\text{ of variable }r,\ i=(l-1)n+r,\\
\frac{(\lambda_{1}\lambda_{2}s_{r})^{2}}{(l^{\lambda_{3}}s_{j})^{2}},\text{ for cross-lag }l\text{ of variable }r\neq j,\ i=(l-1)n+j,
\end{cases}
\end{equation}
see e.g. \citet{karlsson2013}. The prior hyperparameters are refered
to as: \emph{overall-shrinkage }$\lambda_{1}$, \emph{cross-lag shrinkage}
$\lambda_{2}$,\emph{ }and \emph{lag-decay} $\lambda_{3}$\emph{.}

The steady-state formulation of the BVAR makes it non-linear in the
parameters, which complicates the estimation of the model. However,
a simple Gibbs sampling scheme can be used to sample from the posterior
distribution of the model, see \citet{villani2009}. The structure
of the model makes the Gibbs sampling very efficient. However, the
number of parameters in $\Pi$ is $n^{2}p$ so the matrix inversion
for computing full conditional posterior covariance of $\Pi$ requires
$O((n^{2}p)^{3})$ operations in each Gibbs iteration, unless sparsity
is used. This a big bottleneck for large VAR-systems and makes them
unpractical. There has been recent innovations in modeling large-scale
BVARs by reformulating the prior such that the inverse of the covariance
matrix can be obtained for a series of inversions of smaller matrices
\citep{carriero2019large,chan2020large}. Our paper is instead in
the recent VI strand of the literature \citep{koop2018variational,gefang2019variational}
where the posterior is approximated by optimization instead of simulation,
thereby reducing the number of matrix inversions substantially.

\subsection{Variational Inference}

Bayesian inference is based on the posterior distribution of the model
parameters
\begin{equation}
p(\theta|y)=\frac{p(y|\theta)p(\theta)}{\int_{\theta}p(y|\theta)p(\theta)d\theta}\propto p(y|\theta)p(\theta).\label{eq:posterior}
\end{equation}
The \emph{marginal likelihood }in the denominator is usually intractable
for most realistic problems and the posterior is most commonly explored
by Markov chain Monte Carlo (MCMC) simulation where the proportional
form in (\ref{eq:posterior}) is sufficient. MCMC draws converge in
distribution to the target posterior $p(\theta|y)$ and averages of
functions of the simulated parameters converge to posterior expectations.
Even though MCMC is extremely useful and works very well in many applications
it can be computationally expensive, especially when $\theta$ is
high-dimensional. This is a major issue for BVARs with many predictors,
especially since the covariance matrix of the VAR-dynamics has to
be inverted in every iteration.

Variational inference (VI) approximates $p(\theta|y)$ with a simpler
probability density $q(\theta)$ belonging to a tractable family of
distributions, $\mathcal{Q}$. The approximation is formulated as
an optimization problem where the objective is to find the member
of $\mathcal{Q}$ closest to $p(\theta|y)$ in the following Kullback-Leibler
divergence sense \citep{blei2017variational}

\begin{equation}
q^{*}=\underset{q\in\mathcal{Q}}{\arg\min}\:\text{KL}\left(q(\theta)\,||\,p(\theta|y)\right),
\end{equation}
and $q^{*}(\theta)$ is then used an approximation for $p(\theta|y)$.
Note that without restrictions on $\mathcal{Q}$ we will end up approximating
the posterior with itself, which is clearly not useful. The goal is
to consider a family of candidate distributions that are as flexible
as possible, but still provides us with a tractable solution that
is convenient to optimize. A further important thing to note is that
\begin{equation}
\text{KL}\left(q(\theta)||p(\theta|y)\right)=-\int q(\theta)\log\frac{p(\theta|y)p(\theta)}{q(\theta)}d\theta+\log p(y),\label{eq:KLandELBO}
\end{equation}
which means that minimizing the KL divergence is the same as maximizing
\begin{equation}
\int q(\theta)\log\frac{p(\theta|y)p(\theta)}{q(\theta)}d\theta,\label{eq:ELBO}
\end{equation}
since the so called the \emph{evidence} $\log p(y)$ does not depend
on $q(\cdot)$. Since KL is always non-negative the quantity in (\ref{eq:ELBO})
is a lower bound on $\log p(y)$, and therefore often referred to
as the \emph{evidence lower bound (ELBO)}.

There exist a large literature on how to select $\mathcal{Q}$, and
the three main alternatives are \emph{mean-field} (MFVI), \emph{fixed-form
(FFVI) and structured mean-field (SMFVI). }MFVI makes the simplifying
assumption that we may ignore the posterior dependence, i.e. we have
$q(\theta_{1},\theta_{2},\dots,\theta_{k})=q_{1}(\theta_{1})q_{2}(\theta_{2})\dots q_{k}(\theta_{k}).$
This is of course restrictive, but it should be noted that no parametric
assumptions is made on the factors $q_{j}(\theta_{j})$, their functional
forms are determined optimally subject to the independence restriction.
FFVI instead assumes that $q$ comes from a specific class of distributions,
parametrized by a vector of \emph{variational hyperparameters}, $\lambda$,
and minimizes the KL of the posterior from the approximating distribution
$q_{\lambda}$ w.r.t. $\lambda$. It is common to use $q_{\lambda}(\theta)=N(\theta|\mu,\Omega)$
as the approximating variational family, with $\lambda$ consisting
of $\mu$ and the Cholesky factor of $\Omega$. Finally, structured
mean-field is similar to block Gibbs sampling where blocks of parameters
are sampled jointly from their multivariate full conditional posterior
\citep{ormerod2010explaining}. In SMFVI one assumes independent blocks,
but full posterior dependence among the parameters within the blocks,
and optimally chosen functional forms for each block. This is the
form used here as it is perfectly suited for the steady-state BVAR
with three blocks of parameters, $\Pi$, $\Psi$ and $\Sigma$.

\subsection{Structured mean field VI for the steady-state BVAR}

The SMF approximation is obtained by the independence factorization
of the posterior as $q(\theta)=\underset{j=1}{\overset{k}{\prod}}q_{j}(\theta_{j})$,
where $\theta_{j}$ is a block of parameters. The factors $q_{j}(\theta_{j})$
are determined optimally by the maximizing the evidence lower bound
(ELBO) in (\ref{eq:ELBO}). Maximizing the ELBO is done by a coordinate
ascent approach where we cycle through each parameter block and iteratively
maximize the ELBO w.r.t. each $q_{j}^{*}(\theta_{j})$ while fixing
all other $q$:s, see e.g. \citet{bishop2006pattern}. The optimal
solution is given by \citep{bishop2006pattern}:
\begin{equation}
q_{j}^{*}(\theta_{j})=\exp\left\{ \text{E}_{\theta_{-j}}\left[\log p(y,\theta)\right]\right\} +const
\end{equation}
or on the log scale
\begin{equation}
\log q_{j}^{*}(\theta_{j})\propto\text{E}_{\theta_{-j}}\left[\log p(y|\theta)+\log p(\theta)\right],
\end{equation}
where $\text{E}_{\theta_{-j}}$ denotes the expectation with respect
to $\prod_{k\neq j}q_{k}(\theta_{k})$, i.e. the variational factors
of all parameters except the one currently being updated.

Using the factorization $q(\Pi,\Psi,\Sigma)=q_{\Pi}(\Pi)q_{\Psi}(\Psi)q_{\Sigma}(\Sigma)$
with the same three parameter blocks as in the original Gibbs sampler
of \citet{villani2009} we can obtain VI updates from the optimal
solutions

\begin{equation}
\log q_{j}^{*}(\theta_{j})\propto\text{E}_{\theta_{-j}}\left[\log p(y|\Pi,\Psi,\Sigma)+\log p(\Pi|\Psi,\Sigma)+\log p(\Sigma|\Psi)+\log p(\Psi)\right],
\end{equation}
in closed form. The following update steps are derived in Appendix
\ref{sec:Mean-field-derivations} and the corresponding VI algorithm
is given in Algorithm \ref{algVI}.
\begin{itemize}
\item Update step for $\Psi$:
\begin{equation}
q_{\psi}^{*}\left(\psi\right)=N(\psi|\mu_{\psi},\Omega_{\psi})
\end{equation}
with
\[
\begin{aligned}\Omega_{\psi}^{-1}= & \Omega_{\psi|q_{\Pi}q_{\Sigma}}^{-1}+\underline{\Omega}_{\psi}^{-1}\\
\mu_{\psi}= & \Omega_{\psi}\left(m_{\psi|q_{\Pi}q_{\Sigma}}+\underline{\mu}_{\psi}\underline{\Omega}_{\psi}\right).
\end{aligned}
\]
\item Update step for $\Pi$:
\begin{equation}
q_{\Pi}^{*}\left(\text{vec}\Pi\right)=N(\text{vec}\Pi|\mu_{\Pi},\Omega_{\Pi})
\end{equation}
with
\[
\begin{aligned}\Omega_{\Pi}^{-1}= & \Omega_{\Pi|q_{\psi}q_{\Sigma}}^{-1}+\underline{\Omega}_{\Pi}^{-1}\\
\mu_{\Pi}= & \Omega_{\Pi}\left(m_{\Pi|q_{\psi}q_{\Sigma}}+\underline{\mu}_{\pi}\underline{\Omega}_{\Pi}\right).
\end{aligned}
\]
\item Update step for $\Sigma$:
\begin{equation}
q_{\Sigma}^{*}\left(\Sigma\right)=IW(\Sigma|\overline{\nu},\overline{S})
\end{equation}
with
\[
\begin{aligned}\overline{\nu}= & T+\underline{\nu}\\
\overline{S}= & \tilde{S}_{\Sigma|q_{\psi}q_{\Pi}}+\underline{S}.
\end{aligned}
\]
\end{itemize}
Note that underlined letters refers to prior parameters set by the
user and the arguments $m_{\psi|q_{\Pi}q_{\Sigma}},\Omega_{\psi|q_{\Pi}q_{\Sigma}}^{-1},m_{\Pi|q_{\psi}q_{\Sigma}},\Omega_{\Pi|q_{\psi}q_{\Sigma}}^{-1}\text{ and }\tilde{S}_{\Sigma|q_{\psi}q_{\Pi}}$
are recursively updated over during the course of the VI iterations.
Details regarding the VI updating equations can be found in Appendix
\ref{sec:Mean-field-derivations}. 
\begin{algorithm}[h]
\begin{enumerate} 
  \item[\textbf{1}] \textbf{Initialization} 
  \begin{enumerate} 
    \item Select starting values, define the priors and select tolerance level. 
  \end{enumerate} 
  \item[\textbf{2}] \textbf{For} i = 1,2,... while \textbf{criteria}>\textbf{tolerance}
  \begin{enumerate}
    \item Update $\mu_{\Pi}^{(i)}$ and $\Omega_{\Pi}^{(i)}$ in $q^{(i)}_\Pi(\Pi)$ given $q^{(i-1)}_\Sigma$ and $q^{(i-1)}_\psi$. 
    \item Update $\mu_{\Sigma}^{(i)}$ and $\Omega_{\Sigma}^{(i)}$ in $q^{(i)}_\Sigma(\Sigma)$ given $q^{(i)}_\Pi$ and $q^{(i-1)}_\psi$.
    \item Update $\mu_{\psi}^{(i)}$ and $\Omega_{\psi}^{(i)}$ in  $q^{(i)}_\psi(\psi)$ given $q^{(i)}_\Pi$ and $q^{(i)}_\Sigma$.
    \item Update \textbf{criteria}.
  \end{enumerate}
\item[\textbf{3}] \textbf{End while}
\end{enumerate}

\caption{Structured mean field variational Inference\label{algVI}}
\end{algorithm}

After converge, $q^{*}(\theta)$ is a product of the three easily
sampled standard distributions in (11-13). This means that the posterior
for any function $f(\theta)$ of the parameters, e.g. impulse response
functions, are cheaply obtained by direct iid simulation from $q^{*}(\theta)$
after converge and computing $f(\theta)$ for each draw.

\subsection{Log predictive scores}

The \emph{log predictive score }(LPS), see e.g. \citet{geweke2007smoothly}
and \citet{villani2012generalized}, is a commonly used Bayesian model
comparison criteria with the advantage of being much more robust to
prior specification than the marginal likelihood. The LPS used here
is defined as

\begin{equation}
LPS=\sum_{t=s+1}^{T}\log\int p(y_{t+1}|y_{1:t},\theta)p(\theta\vert y_{1:t})d\theta,\label{eq:lps}
\end{equation}
where $s$ is a number of observations used to train the model; we
set $s=30$ for the rest of the paper. The LPS is often used in place
of the marginal likelihood when calculating posterior model probabilities
to increase robustness with respect the prior specification. 

Using draws from $\theta^{(i)}\sim p(\theta\vert y_{1:t})$, the LPS
can be estimated by

\begin{equation}
\widehat{LPS}=\sum_{t=s+1}^{T}\log\left[\frac{1}{N}\sum_{n=1}^{N}p(y_{t+1}|y_{1:t},\theta^{(i)})\right].\text{\textasciiacute}\label{eq:LPShat}
\end{equation}
Note that computing the LPS by Gibbs sampling is very costly since
we need to run a complete Gibbs sampling run for each of the intermediate
posteriors $p(\theta\vert y_{1:t})$ for $t=s+1,\ldots,T$. It is
possible to use reweighted draws from the final posterior $p(\theta\vert y_{1:T})$
in an importance sampling approximation \citep{geweke1999using},
but this needs careful monitoring of the importance weights and is
rarely done in practice. VI can instead draw the $\theta^{(i)}$ in
(\ref{eq:LPShat}) by fast direct simulation from the VI approximation
of each intermediate posterior $q(\Pi,\Psi,\Sigma\vert y_{1:t})=q_{\Pi}(\Pi\vert y_{1:t})q_{\Psi}(\Psi\vert y_{1:t})q_{\Sigma}(\Sigma\vert y_{1:t})$
for each $t=s+1,\ldots,T$. The optimization of each $q(\Pi,\Psi,\Sigma\vert y_{1:t})$
is extremely quick since the optimal VI hyperparameters from the previous
time step at $t-1$ are typically excellent intial values \citep{nott2012regression}.

\section{Simulation Experiments}

In this section, we compare the VI approximation to the standard Gibbs
sampling approach, which is considered to be the ground truth. We
are particularly interested in demonstrating the performance of VI
for: i) different degrees of persistence of the time series system
and ii) different informativeness of the steady-state prior. \citet{villani2009}
points out that inference about the steady-state is increasingly difficult,
and the Gibbs sampler become very inefficient, as the process approaches
one or more unit roots \emph{if} a non-informative vague prior is
used; there is local non-identification in the sense that $\Psi$
becomes less and less identified the closer $\Pi$ is to the non-stationary
region; see \citet{villani2006inference} for more details. This has
not concerned practitioners since the whole point of using the steady-state
BVAR is that one has relatively strong prior information about the
steady-state and the posterior of $\Psi$ will be dominated by the
prior whenever $\Pi$ is close to the non-stationary region. 

The local non-identification implies posterior dependence between
$\Pi$ and $\Psi$, so VI is expected too work less well here, at
least when a noninformative prior is used. So it is an interesting
setup to explore the limitations of VI, even if this setting is rarely
used in practice. 

We will first compare VI and Gibbs on moderately persistent data with
and without strong prior beliefs on the steady-state, and subsequently
move on to the more challenging case with a highly persistent process.
We compare parameter posteriors, predictive distributions at different
forecast horizons, as well as LPS for different lag lengths. Throughout
the whole section we use the common hyperparameter set-up of $\lambda_{1}=0.2,\lambda_{2}=0.5,\lambda_{3}=1$.
In all illustrations, blue lines represent the VI approach and red
lines the Gibbs approach.

\subsection{Moderate persistence}

We first simulate a dataset of $T=100$ observations from the following
moderately persistent VAR(1) model:

\begin{equation}
\begin{aligned}y_{t}-\Psi= & \Pi(y_{t-1}-\Psi)+\varepsilon_{t},\end{aligned}
\qquad\text{where }\varepsilon_{t}\sim N(0,\Sigma),
\end{equation}
\begin{equation}
\begin{aligned}\Pi=\begin{pmatrix}0.45 & 0.5\\
0.1 & 0.65
\end{pmatrix}, & \:\Psi=\begin{pmatrix}6\\
1
\end{pmatrix}, & \Sigma=\begin{pmatrix}1 & 0.4\\
0.4 & 1
\end{pmatrix}\end{aligned}
,
\end{equation}
with the eigenvalues of $\Pi$ being $0.31$ and 0.79. The Gibbs sampling
is run with $100\,000$ posterior draws where $20\,000$ draws are
used as a burn-in sample, while the VI updates are run until convergence,
i.e.~until the posterior parameters does not change anymore. We use
both a weak and an informative prior on the steady-state. The weak
prior is $N(\underline{\Psi},\underline{\Omega}_{\Psi})$ with $\underline{\Psi}=\Psi$
and $\underline{\Omega}_{\Psi}=\text{diag}(100,100)$, which is basically
flat around the true parameter values. The informative prior has the
same mean, but uses $\underline{\Omega}_{\Psi}=\text{diag}(0.25,0.25)$,
hence giving $95\%$ intervals of approximately $\pm1$ around the
mean, which are quite common in applications, see Section \ref{subsec:Real_data}.

\begin{figure}[H]
\centering{}\includegraphics[scale=0.35]{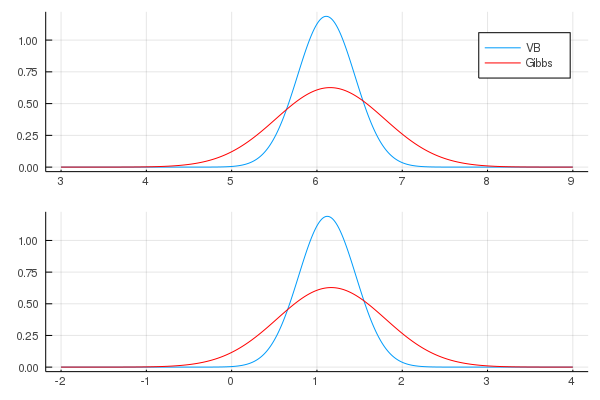}\includegraphics[scale=0.35]{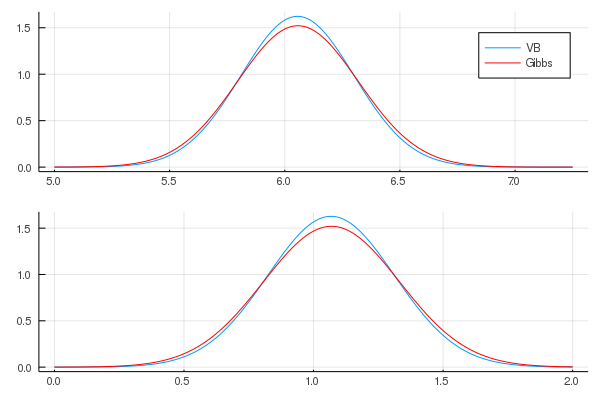}\caption{Moderate persistent VAR. Posterior distribution for $\Psi$ an informative
prior (left column) and a weak prior (right columns) on the steady-states.\label{fig:low_persist_tight_prior_psi}}
\end{figure}

Figure \ref{fig:low_persist_tight_prior_psi} shows the Gibbs and
VI posteriors on $\Psi$ for both the weak (left column) and the informative
(right column) prior. The VI posterior gets the posterior location
right when using a weak prior, but underestimates the posterior variance;
this is a common situation for mean-field VI and comes from the assumed
independence between parameter blocks. The right column of Figure
\ref{fig:low_persist_tight_prior_psi} shows that VI has much better
accuracy when an more informative, and realistic, prior is used. 

\begin{figure}
\includegraphics[scale=0.36]{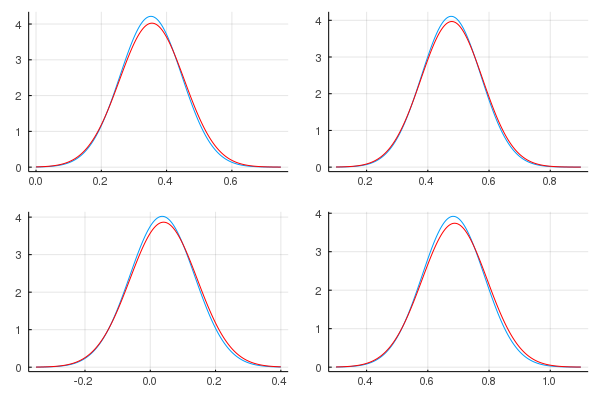}\includegraphics[scale=0.36]{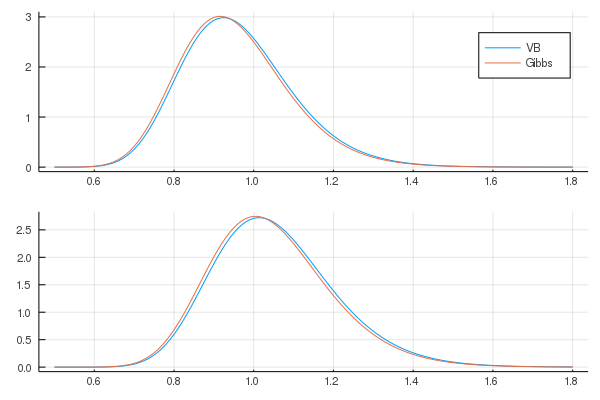}\caption{Moderate persistent VAR. Marginal posterior distributions for $\Pi$
and $\Sigma$ for a weak prior on the steady-states.\label{fig:low_persist_PI_Sigma}}
\end{figure}

Figure \ref{fig:low_persist_PI_Sigma} shows that the VI approximations
for $\Pi$ and $\Sigma$ are close to indistinguishable from the Gibbs
sampling posteriors even when using a weak prior on the steady-states.
With the informative prior, the VI and Gibbs posteriors are extremely
similar and are therefore not shown.

\begin{figure}[H]
\centering{}\includegraphics[scale=0.52]{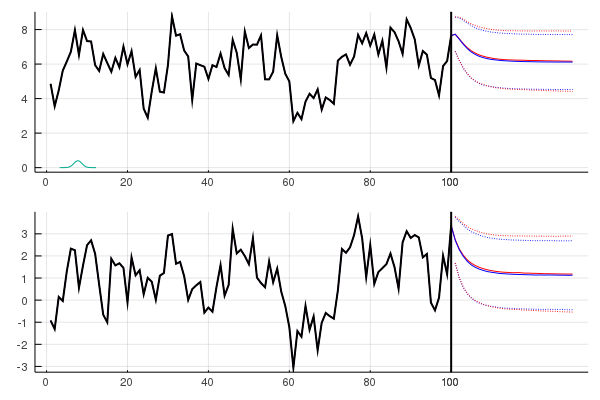}\caption{$h$-steps-ahead forecast distributions (mean and one std deviation
bands), $h=1,\ldots,30$, using an uninformative prior on the steady-states
for the moderately persistent VAR. Red lines show the results from
Gibbs sampling and blue lines for VI.\label{fig:weak_Low_persistent_forecast}}
\end{figure}

Figure \ref{fig:weak_Low_persistent_forecast} shows that the point
forecasts produced from Gibbs and VI under the weak prior are virtually
identical, and the forecast intervals are only slightly distorted
by VI. With an informative prior the forecasts and the intervals are
visually identical and not shown here.

Table \ref{tab:model_prob} compares the approximations of posterior
model probabilities for different lags for the VAR computed by normalizing
the LPS. The results are very close even for the weak prior.

\begin{table}[H]
\centering{}%
\begin{tabular}{lccccccc}
\hline 
 & \multicolumn{3}{c}{Gibbs sampling} &  & \multicolumn{3}{c}{VI}\tabularnewline
Number of lags: & $1$ & $2$ & $3$ &  & $1$ & $2$ & $3$\tabularnewline
\cline{2-4} \cline{3-4} \cline{4-4} \cline{6-8} \cline{7-8} \cline{8-8} 
Weak prior & $0.62$ & $0.23$ & $0.15$ &  & $0.64$ & $0.20$ & $0.16$\tabularnewline
Informative prior & $0.65$ & $0.19$ & $0.16$ &  & $0.67$ & $0.18$ & $0.15$\tabularnewline
\hline 
\end{tabular}\caption{LPS-based posterior model probabilities for lag length for the moderately
persistent VAR.\label{tab:model_prob}}
\end{table}

In summary, for a moderately persistent VAR with the kind of prior
used in practical work, the results from VI are essentially the same
as the ones from Gibbs sampling. When a very weak prior is used on
the steady-state, the posterior variance for $\Psi$ is substantially
underestimated, but all other aspects, including predictive distributions
and LPS are still very close to the results from Gibbs sampling.

\subsection{High persistence}

We now consider a much more challenging situation with higher persistence
by replacing the two diagonal elements of the VAR dynamics with $0.6$
and $0.8$, respectively. All other parameters, as well as prior settings,
remain the same. The eigenvalues of the companion matrix are now $0.96$
and $0.46$, hence very close to a unit root. We again note that one
would typically not use the steady-state BVAR with such an uninformative
prior in this setting, but the exercise is useful as an extreme case.

\begin{figure}[b]
\centering{}\includegraphics[scale=0.35]{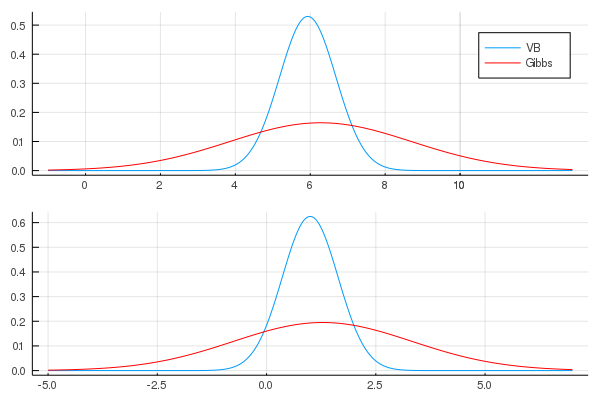}\includegraphics[scale=0.35]{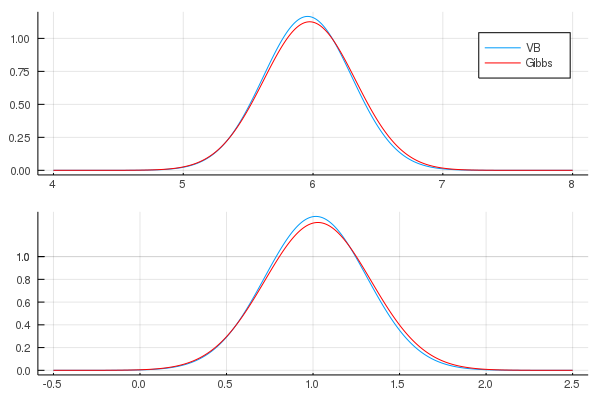}\caption{Highly persistent VAR. Posterior distribution for $\Psi$ using a
weak prior (left column) and a informative prior (right columns) on
the steady-states.\label{fig:persist_psi}}
\end{figure}

\begin{figure}
\includegraphics[scale=0.35]{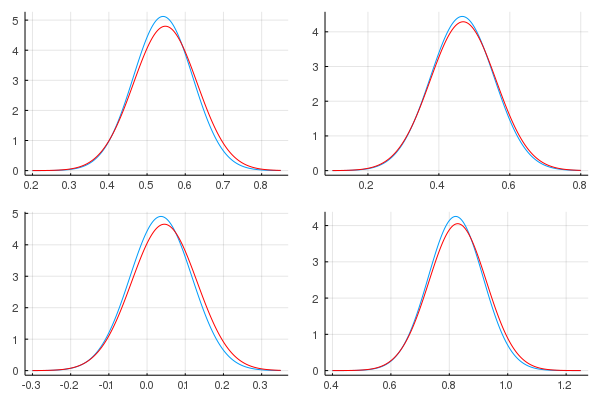}\includegraphics[scale=0.35]{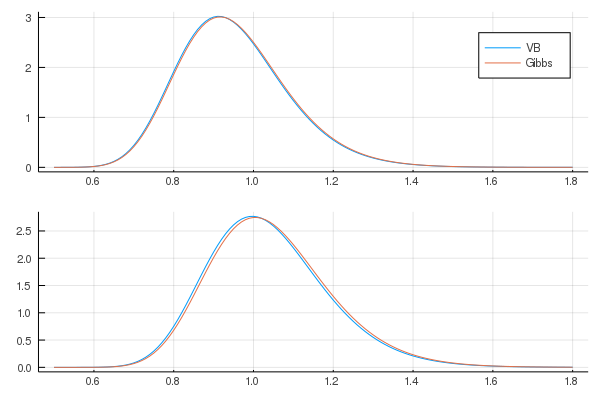}\caption{High persistent VAR. Marginal posterior distributions for $\Pi$ and
$\Sigma$ for a weak prior on the steady-states.\label{fig:high_persist_PI_Sigma}}
\end{figure}

The left column of Figure \ref{fig:persist_psi} shows that the location
of the VI posterior distribution for $\Psi$ is still accurate, but
the variance is now severely underestimated compared to the Gibbs
sampler. Figure \ref{fig:high_persist_PI_Sigma} shows that the posterior
approximation for $\Pi$ and $\Sigma$ are nevertheless excellent
even for the uninformative prior.

\begin{figure}
\centering{}\includegraphics[scale=0.55]{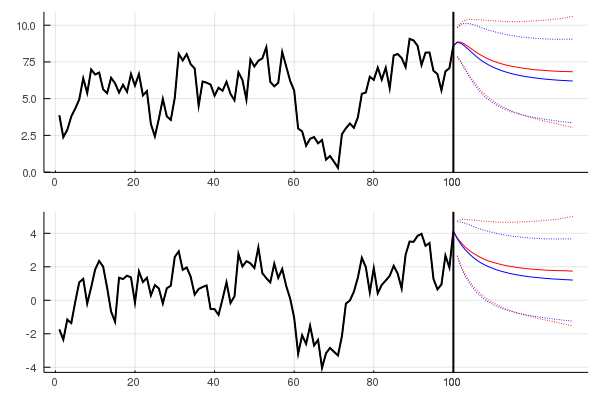}\caption{$h$-steps-ahead forecast distributions (mean and one std deviation
bands), $h=1,\ldots,30$, using an uninformative prior on the steady-states
for the highly persistent VAR. Red lines show the results from Gibbs
sampling and blue lines for VI. \label{fig:weak_persist_forecast}}
\end{figure}

\begin{figure}
\centering{}\includegraphics[scale=0.55]{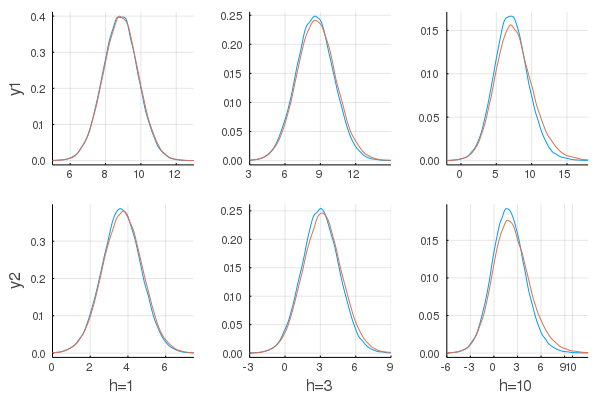}\caption{Out-of-sample forecast densities on 1, 3, and 10 steps-ahead predictions
for the high persistent series using an uninformative prior on the
steady-states. \label{fig:weak_persist_forecast_dist}}
\end{figure}

\begin{figure}
\centering{}\includegraphics[scale=0.55]{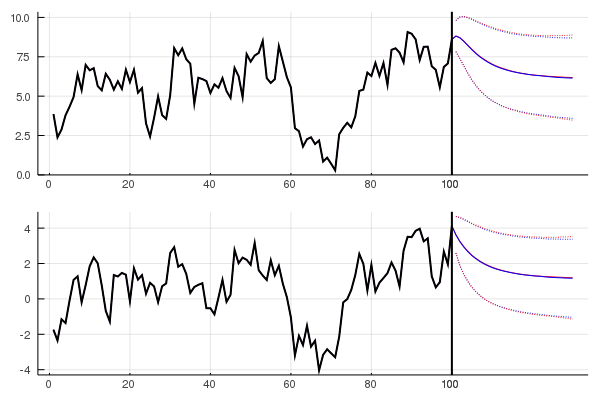}\caption{$h$-steps-ahead forecast distributions (mean and one std deviation
bands), $h=1,\ldots,30$, using an informative prior on the steady-states
for the highly persistent VAR. Red lines show the results from Gibbs
sampling and blue lines for VI.\label{fig:tight_persistent_forecast}}
\end{figure}

Figure \ref{fig:weak_persist_forecast} shows that the inaccurate
VI posterior for $\Psi$ for the noninformative prior leads to somewhat
inaccurate mean predictions and prediction intervals, especially at
longer horizons where VI also underestimates the forecasting uncertainty.
Figure \ref{fig:weak_persist_forecast_dist} shows however that the
overall accuracy of the marginal predictive distributions are not
as bad as one would think from the intervals in Figure \ref{fig:weak_persist_forecast}.
More, importantly, Figure \ref{fig:tight_persistent_forecast} shows
that for the more realistic informative prior, the predictive mean
and intervals from VI are indistinguishable from those of Gibbs sampling;
the same is true for the whole predictive densities (not shown).

Table \ref{tab:model_prob-persist} shows that the LPS approximation
from VI are excellent for both priors. 

\begin{table}[H]
\centering{}%
\begin{tabular}{lccccccc}
\hline 
 & \multicolumn{3}{c}{Gibbs-sampling} &  & \multicolumn{3}{c}{VI}\tabularnewline
Number of lags: & $1$ & $2$ & $3$ &  & $1$ & $2$ & $3$\tabularnewline
\cline{2-4} \cline{3-4} \cline{4-4} \cline{6-8} \cline{7-8} \cline{8-8} 
Weak & $0.72$ & $0.16$ & $0.12$ &  & $0.77$ & $0.13$ & $0.10$\tabularnewline
Informative & $0.73$ & $0.15$ & $0.12$ &  & $0.77$ & $0.13$ & $0.10$\tabularnewline
\hline 
\end{tabular}\caption{LPS-based posterior model probabilities for lag length for the strongly
persistent VAR.\label{tab:model_prob-persist}}
\end{table}

\begin{figure}[H]
\centering{}\includegraphics[scale=0.36]{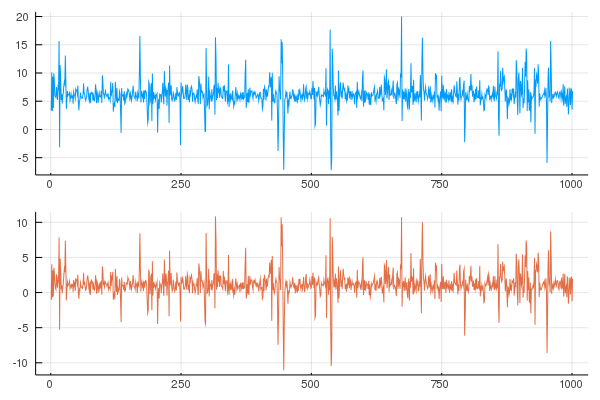}\includegraphics[scale=0.36]{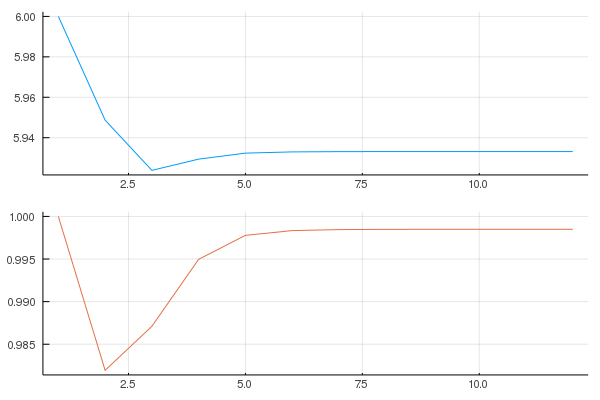}\caption{Last 1000 Gibbs draws (left) and the 12 first VI iterations for the
steady-state parameters in the case of a highly persistent time series
and an uninformative prior on the steady-states.\label{fig:mcmc_conv}}
\end{figure}

The main advantage of VI is its speed. A time benchmarking exercise
will be provided in the real data study in Subsection \ref{subsec:Real_data}
where we can see that VI scales very well compared to Gibbs sampling.
Here just note that the Gibbs sampler may mix poorly when the process
is strongly persistent and a noninformative prior is used. The left
hand side of Figure \ref{fig:mcmc_conv} shows that the Gibbs sampler
enters periods of high volatility for the steady-states; this happens
when the $\Pi$ draws are close to the non-stationary region \citep{villani2009}.
The right part of Figure \ref{fig:mcmc_conv} shows that the VI algorithm
needs only a very small number of iterations to converge. 

\subsection{Application to US macroeconomic data\label{subsec:Real_data}}

\subsection*{Data and priors}

To get a sense of how well VI performs in practice we use a real data
study with 23 macroeconomic time series from the FRED database. The
same data set is used in e.g. \citet{giannone2015} and \citet{gustafsson2020bayesian}
and is divided into a medium-sized model containing seven variables
as indicated in Table \ref{tab:data} and a large-sized model where
all data is used except for\emph{ real investments,} which is excluded
from the large BVAR since both \emph{residential-} and \emph{non-residential
investments} are included. All series are analyzed on a quarterly
frequency and are made stationary, to be in line with the prior assumption
of a steady-state. The priors for the steady states and the transformations
of the data are the same as in \citet{gustafsson2020bayesian} and
can be found in Table \ref{tab:data}. The prior mean on first own
lag of the \emph{FED interest} rate and the \emph{GDP-deflator }is
set to 0.6 to reflect some degree of persistence and the prior mean
for rest of the VAR-dynamics is set to zero.

\begin{table}
\begin{centering}
\begin{tabular}{lccccc}
\multicolumn{6}{c}{{\footnotesize{}Variable names and transformations}}\tabularnewline
\hline 
\multirow{2}{*}{{\footnotesize{}Variables}} & {\footnotesize{}Mnemonic} & \multirow{2}{*}{{\footnotesize{}Transform}} & \multirow{2}{*}{{\footnotesize{}Medium}} & \multirow{2}{*}{{\footnotesize{}Freq.}} & \multirow{2}{*}{{\footnotesize{}Prior}}\tabularnewline
 & {\footnotesize{}(FRED)} &  &  &  & \tabularnewline
\hline 
\hline 
{\footnotesize{}Real GDP} & {\footnotesize{}GDPC1} & {\footnotesize{}$400\times\text{diff-log}$} & {\footnotesize{}x} & {\footnotesize{}Q} & {\footnotesize{}(2.5;3.5)}\tabularnewline
{\footnotesize{}GDP deflator} & {\footnotesize{}GDPCTPI} & {\footnotesize{}$400\times\text{diff-log}$} & {\footnotesize{}x} & {\footnotesize{}Q} & {\footnotesize{}(1.5;2.5)}\tabularnewline
{\footnotesize{}Fed funds rate} & {\footnotesize{}FEDFUNDS} & {\footnotesize{}-} & {\footnotesize{}x} & {\footnotesize{}Q} & {\footnotesize{}(4.3,5.7)}\tabularnewline
{\footnotesize{}Consumer price index} & {\footnotesize{}CPIAUCSL} & {\footnotesize{}$400\times\text{diff-log}$} &  & {\footnotesize{}M} & {\footnotesize{}(1.5;2.5)}\tabularnewline
{\footnotesize{}Commodity prices} & {\footnotesize{}PPIACO} & {\footnotesize{}$400\times\text{diff-log}$} &  & {\footnotesize{}Q} & {\footnotesize{}(1.5;2.5)}\tabularnewline
{\footnotesize{}Industrial production} & {\footnotesize{}INDPRO} & {\footnotesize{}$400\times\text{diff-log}$} &  & {\footnotesize{}Q} & {\footnotesize{}(2.3;3.7)}\tabularnewline
{\footnotesize{}Employment} & {\footnotesize{}PAYEMS} & {\footnotesize{}$400\times\text{diff-log}$} &  & {\footnotesize{}Q} & {\footnotesize{}(1.5;2.5)}\tabularnewline
{\footnotesize{}Employment, service sector} & {\footnotesize{}SRVPRD} & {\footnotesize{}$400\times\text{diff-log}$} &  & {\footnotesize{}Q} & {\footnotesize{}(2.5;3.5)}\tabularnewline
{\footnotesize{}Real consumption} & {\footnotesize{}PCECC96} & {\footnotesize{}$400\times\text{diff-log}$} & {\footnotesize{}x} & {\footnotesize{}Q} & {\footnotesize{}(2.3;3.7)}\tabularnewline
{\footnotesize{}Real investment} & {\footnotesize{}GPDIC1} & {\footnotesize{}$400\times\text{diff-log}$} & {\footnotesize{}x} & {\footnotesize{}Q} & {\footnotesize{}(1.5;4.5)}\tabularnewline
{\footnotesize{}Real residential investment} & {\footnotesize{}PRFIx} & {\footnotesize{}$400\times\text{diff-log}$} &  & {\footnotesize{}Q} & {\footnotesize{}(1.5;4.5)}\tabularnewline
{\footnotesize{}Nonresidential investment} & {\footnotesize{}PNFIx} & {\footnotesize{}$400\times\text{diff-log}$} &  & {\footnotesize{}Q} & {\footnotesize{}(1.5;4.5)}\tabularnewline
{\footnotesize{}Personal consumption} & \multirow{2}{*}{{\footnotesize{}PCECTPI}} & \multirow{2}{*}{{\footnotesize{}$400\times\text{diff-log}$}} & \multirow{2}{*}{} & \multirow{2}{*}{{\footnotesize{}Q}} & \multirow{2}{*}{{\footnotesize{}(1.5;4.5)}}\tabularnewline
{\footnotesize{}expenditure, price index} &  &  &  &  & \tabularnewline
{\footnotesize{}Gross private domestic} & \multirow{2}{*}{{\footnotesize{}GPDICTPI}} & \multirow{2}{*}{{\footnotesize{}$400\times\text{diff-log}$}} & \multirow{2}{*}{} & \multirow{2}{*}{{\footnotesize{}Q}} & \multirow{2}{*}{{\footnotesize{}(1.5;4.5)}}\tabularnewline
{\footnotesize{}investment, price index} &  &  &  &  & \tabularnewline
{\footnotesize{}Capacity utilization} & {\footnotesize{}TCU} & {\footnotesize{}-} &  & {\footnotesize{}Q} & {\footnotesize{}(79.3;80.7)}\tabularnewline
{\footnotesize{}Consumer expectations} & {\footnotesize{}UMCSENTx} & {\footnotesize{}diff} &  & {\footnotesize{}Q} & {\footnotesize{}(-0.5, 0.5)}\tabularnewline
{\footnotesize{}Hours worked} & {\footnotesize{}HOANBS} & {\footnotesize{}$400\times\text{diff-log}$} & {\footnotesize{}x} & {\footnotesize{}Q} & {\footnotesize{}(2.5;3.5)}\tabularnewline
{\footnotesize{}Real compensation/hour} & {\footnotesize{}AHETPIx} & {\footnotesize{}$400\times\text{diff-log}$} & {\footnotesize{}x} & {\footnotesize{}Q} & {\footnotesize{}(1.5;2.5)}\tabularnewline
{\footnotesize{}One year bond rate} & {\footnotesize{}GS1} & {\footnotesize{}diff} &  & {\footnotesize{}Q} & {\footnotesize{}(-0.5;0.5)}\tabularnewline
{\footnotesize{}Five years bond rate} & {\footnotesize{}GS5} & {\footnotesize{}diff} &  & {\footnotesize{}M} & {\footnotesize{}(-0.5,0.5)}\tabularnewline
{\footnotesize{}SP 500} & {\footnotesize{}S\&P 500} & {\footnotesize{}$400\times\text{diff-log}$} &  & {\footnotesize{}Q} & {\footnotesize{}(-2,2)}\tabularnewline
{\footnotesize{}Effective exchange rate} & {\footnotesize{}TWEXMMTH} & {\footnotesize{}$400\times\text{diff-log}$} &  & {\footnotesize{}Q} & {\footnotesize{}(-1;1)}\tabularnewline
{\footnotesize{}M2} & {\footnotesize{}M2REAL} & {\footnotesize{}$400\times\text{diff-log}$} &  & {\footnotesize{}Q} & {\footnotesize{}(5.5;6.5)}\tabularnewline
\hline 
\end{tabular}\caption{Data Description\label{tab:data}}
\par\end{centering}
\raggedright{}{\footnotesize{}The table shows the 23 US macroeconomic
time series from the FRED database. The column named Prior contains
the steady-state mean $\pm$ one standard deviation.}{\footnotesize\par}
\end{table}

\subsection*{Medium size model}

We use the prior hyperparameters found in \citet{gustafsson2020bayesian}
by the Bayesian optimization with optimized precision (BOOP) algorithm:
$\lambda_{1}=0.27,\lambda_{2}=0.43$ and $\lambda_{3}=0.76$. The
Gibbs sampler is run with $100\,000$ MCMC draws with $20\,000$ draws
used as a burn-in.

\begin{figure}[H]
\centering{}\includegraphics[scale=0.46]{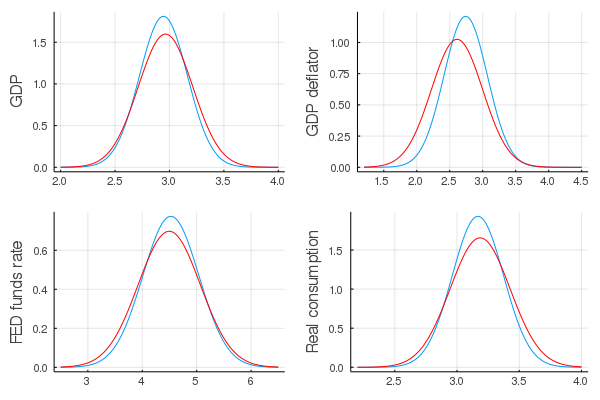}\caption{Posterior distribution for $\Psi$ in the 7-variable VAR; blue and
red lines represents VI and Gibbs sampling, respectively.\label{fig:real_psi}}
\vspace{-0.5cm}
\end{figure}

Figure \ref{fig:real_psi} shows that VI somewhat underestimates the
posterior variance of $\Psi$, but the overall accuracy is quite acceptable.
The steady-state posteriors for the remaining time series are found
in Figure \ref{fig:real_psi-1} in Appendix \ref{sec:List-of-figures}.
The VI approximation for $\Pi$ and $\Sigma$ are nearly perfect,
see Figure \ref{fig:real_Pi} in Appendix \ref{sec:List-of-figures}.
 However, what really matters in practice is how well VI approximates
predictions and more interesting quantities from the model, such as
impulse responses. Since impulse responses are only functions of $\Pi$
and $\Sigma$, their implied VI posterior will also be very accurate.
Figure \ref{fig:real_forecast} and Figure \ref{fig:real_forecast_append}
in the Appendix show that the 1-12 steps-ahead mean forecasts and
intervals for VI are nearly indistinguishable from their Gibbs sampling
counterparts. The same is true for the whole forecast distribution
(not shown).
\begin{figure}[H]
\centering{}\includegraphics[scale=0.5]{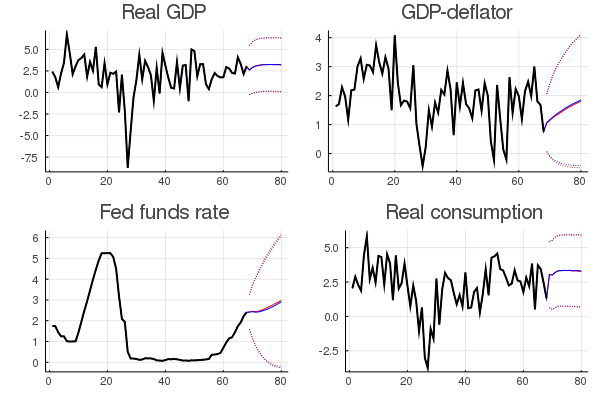}\caption{Out of sample forecasts for the medium sized model. Solid blue line
represent point-forecasts produced by VI, and dotted blue lines are
1 standard deviation posterior predictive intervals. Corresponding
red lines are from Gibbs sampling. \label{fig:real_forecast}}
\end{figure}

To investigate the model selection properties of the VI approximation
on the real data set we again calculate posterior model probabilities
for several model alternatives via the LPS. In this exercise we treat
the number of lags as fixed and investigate the predictive behavior
when changing the prior hyperparameter $\lambda_{1}$. We let our
hypothesized main alternative be the hyperparameter setup in \citet{gustafsson2020bayesian}
with $\lambda_{1}=0.27$, and compare it to the alternative settings:
$(i):\lambda_{1}=0.1$, $(ii):\lambda_{1}=0.2$, $(iii):\lambda_{1}=0.4$,
$(iv):\lambda_{1}=0.5$, and $(v):\lambda_{1}=10$. Table \ref{tab:Bayes_factor-1}
shows that the estimated model probabilities differ a little between
approaches, but both clearly identify $\lambda_{1}=0.27$ as the best
value for the hyperparameter.

\begin{table}[H]
\centering{}%
\begin{tabular}{ccccccc}
\hline 
$\lambda_{1}$ & $0.1$ & $0.2$ & $0.27$ & $0.4$ & $0.5$ & $10$\tabularnewline
\hline 
Variational Bayes & $0.00..$ & $0.08$ & $0.84$ & $0.08$ & $0.00..$ & $0.00..$\tabularnewline
Gibbs sampling & $0.00..$ & $0.04$ & $0.66$ & $0.20$ & $0.10$ & $0.00..$\tabularnewline
\hline 
\end{tabular}\caption{Posterior model probabilities for different hyperparameter settings.\label{tab:Bayes_factor-1}}
\end{table}

To get a sense of how fast VI computes the LPS we consider the computation
times in a single run from Table \ref{tab:Bayes_factor-1}. As before
the Gibbs sampler is used with $100\,000$ MCMC draws with a burn-in
of $20\,000$, while VI iterates until convergence and then takes
$80\,000$ draws by direct simulations from the VI-approximation.
The time series length for the medium-sized model is $218$, and since
the first $30$ observations are used as a training sample, both approaches
has to be re-estimated 188 times. In this setting VI required 226
seconds to complete whereas Gibbs sampling took more than $40\,000$
seconds. 

\begin{figure}[H]
\centering{}\includegraphics[scale=0.55]{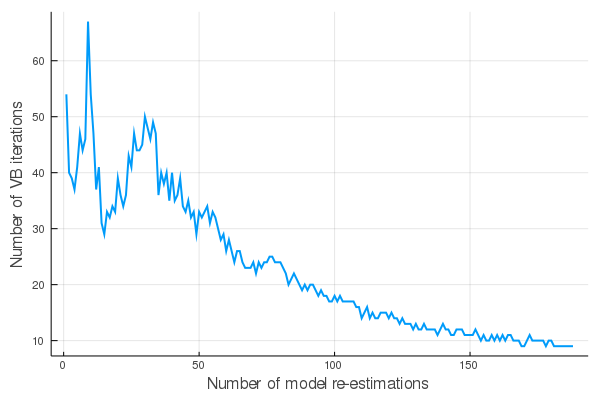}\caption{VI needs fewer iterations to converge when approximating $p(\theta\vert y_{1:t})$
for larger $t$ in the LPS. \label{fig:VI_iterations_over_time}}
\end{figure}

Part of the time gain comes from VIs ability to use of the optimized
variational hyperparameters from the previous time step as initial
values. The initial values improve for later terms in the LPS, as
illustrated in Figure \ref{fig:VI_iterations_over_time}.

\begin{figure}[H]
\centering{}\includegraphics[scale=0.36]{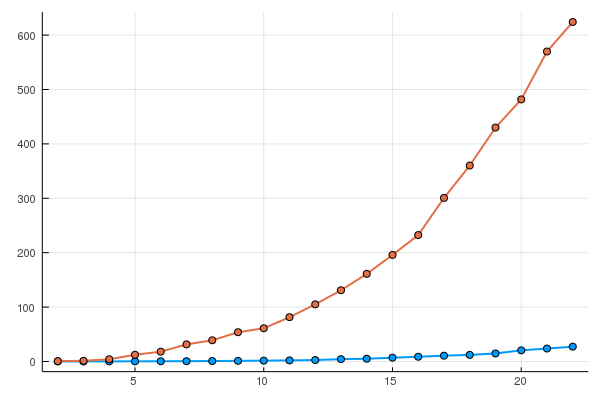}\includegraphics[scale=0.36]{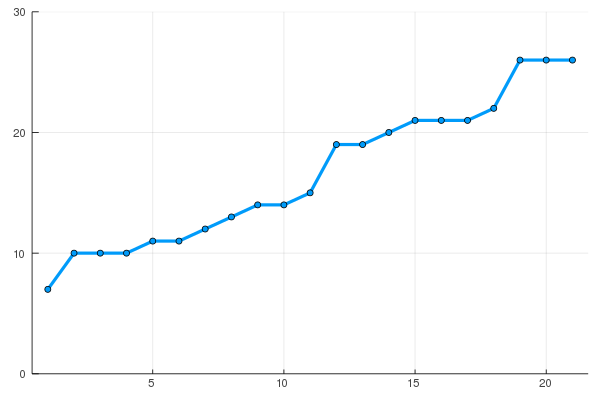}\caption{Computating times in seconds for the LPS with different number of
time series in the VAR-system (left). The number of VI iterations
until convergence as a function of the number of time series (right).\label{fig:VB_speed}}
\end{figure}

To see how well VI scales compared to Gibbs sampling we compare the
computation times for VAR systems of increasing size. We consider
the large data set but start with a subset of only two time series
and subsequently add one time series at a time until all time series
are included in the system. The exercise is carried out with $10\,000$
MCMC draws while the VI iterates until convergence. We can see from
the left side of Figure \ref{fig:VB_speed} that the computational
gains from using the VI approximation are huge, and that VI scales
much better than Gibbs sampling to larger systems. One should note
that the data set that we refer to as ``Large'' is in fact much
smaller than in e.g. \citet{banbura} and \citet{koop2013} where
more than a hundred time series are used and we have not tested the
VI algorithm under those settings yet. The right part of Figure \ref{fig:VB_speed}
shows that the required number of VI iterations increases fairly linearly
in the number of time series.

Similarly, as in the medium-sized model, the posterior distributions
for the steady-states are a little bit off, while the VI approximation
for the other parameter blocks are very accurate (not shown). The
VI approximation of the predictive distributions are highly accurate
as can be seen in Figure \ref{fig:Large_forecasts} in the appendix.

\section{Discussion\label{sec:Discussion}}

We propose a structured mean field variational inference approach
to approximate the parameter posterior and the predictive distribution
for the steady-state BVAR of \citet{villani2009}. The approximation
is very fast compared to the widely used Gibbs sampler and produces
accurate posterior distributions and forecast distributions that are
virtually identical to those from the much more time-consuming Gibbs
sampler.

We also show that the VI approximation can be used to very efficiently
and accurately compute log predictive scores (LPS) for robust Bayesian
model comparison. LPS requires re-estimation of the model for each
time-period and since VI rely on optimization it can use the optimized
variational hyperparameters from the previous time step as excellent
starting values for quick convergence.

The structured mean-field approximation used here assumes independence
between the three parameter blocks. This assumption can be relaxed
using a fixed-form VI strategy at the cost of a slower and less robust
VI algorithm since the VI updates would then no longer be in closed
form. Extensions of the proposed VI algorithm to time-varying latent
steady-steady states and stochastic volatility are in principle straightforward
by adding VI updating steps, but the details needs to be worked out,
and is a interesting future research agenda.

\bibliographystyle{agsm}
\bibliography{ref}

\appendix

\part*{Appendix}

\section{Derivations of the VI updates \label{sec:Mean-field-derivations}}

\subsection*{Preliminary steps}

The structured mean field approximation for the posterior of $\theta=(\Psi,\Pi,\Sigma)$
is given by

\begin{equation}
q(\theta)=q(\Psi,\Pi,\Sigma)=q_{\Psi}(\Psi)q_{\Pi}(\Pi)q_{\Sigma}(\Sigma)
\end{equation}
with optimal approximating densities obtained from

\begin{equation}
\begin{aligned}\log q_{j}(\theta_{j})\propto\, & \text{E}_{\theta_{-j}}\left[\log p(y|\Psi,\Pi,\Sigma)+\log p(\Pi|\Sigma,\Psi)+\log p(\Sigma|\Psi)+\log p(\Psi)\right].\end{aligned}
\end{equation}

For the steady-state BVAR in Section \ref{subsec:Steady-state-BVARs},
we have
\begin{equation}
\begin{aligned}\log q_{j}(\theta_{j})\propto\, & E_{\theta_{-j}}\left[-\frac{nT}{2}\log(2\pi)-\frac{T}{2}\log|\Sigma|-\frac{1}{2}\sum\left(\left(\Pi(L)(\mathbf{y}_{t}-\Psi)\right)^{T}\Sigma^{-1}\Pi(L)(\mathbf{y}_{t}-\Psi)\right)\right.\\
 & -\frac{n}{2}\log(2\pi)-\frac{1}{2}\log|\underline{\Omega}_{\Psi}|-\frac{1}{2}\left(\Psi-\underline{\Psi}\right)^{T}\underline{\Omega}_{\Psi}^{-1}\left(\Psi-\underline{\Psi}\right)\\
 & -\frac{n^{2}p}{2}\log(2\pi)-\frac{1}{2}\log|\underline{\Omega}_{\Pi}|-\frac{1}{2}\left(\Pi-\underline{\Pi}\right)^{T}\underline{\Omega}_{\Pi}^{-1}\left(\Pi-\underline{\Pi}\right)\\
 & \left.-\frac{\underline{\nu}n}{2}\log2+\frac{\underline{\nu}}{2}\log|\underline{S}|-\Gamma\left(\frac{\underline{\nu}}{2}\right)-\frac{\underline{\nu}+n+1}{2}\log|\Sigma|-\frac{1}{2}\text{tr}\left(\underline{S}\Sigma^{-1}\right)\right].
\end{aligned}
\end{equation}

If we ignore terms that are constant w.r.t. to $\theta$ we obtain

\begin{equation}
\begin{aligned}\log q_{j}(\theta_{j})\propto & -\frac{1}{2}E_{\theta_{-j}}\left[T\log|\Sigma|+\sum_{t=1}^{T}\left(\Pi(L)\left(y_{t}-\Psi\right)\right)^{T}\Sigma^{-1}\Pi(L)\left(y_{t}-\Psi\right)\right.\\
 & +\left(\Psi-\underline{\Psi}\right)^{T}\underline{\Omega}_{\Psi}^{-1}\left(\Psi-\underline{\Psi}\right)+\left(\Pi-\underline{\Pi}\right)^{T}\underline{\Omega}_{\Pi}^{-1}\left(\Pi-\underline{\Pi}\right)\\
 & \left.+\left(\underline{\nu}+n+1\right)\log|\Sigma|+\text{tr}\left(\underline{S}\Sigma^{-1}\right)\right]
\end{aligned}
\label{eq:VI_all}
\end{equation}

It will be convenient in the derivations to write the model in vectorized
form

\begin{equation}
y_{\Psi}=(I\otimes X_{\Psi})\pi+e,\qquad\text{where }e\sim N(0,\Sigma\otimes I),
\end{equation}
and $y_{\Psi}=\text{vec}(Y)-\text{vec}(\Psi_{y})$, $X_{\Psi}=X-\Psi_{x}$,
$\pi=\text{vec}(\Pi),$ $\Psi_{y}=\mathbf{1}_{T}\psi^{T}I_{n}=\left(I_{n}\otimes\mathbf{1}_{T}\right)\text{vec}\psi^{T}=V\psi$
and $\Psi_{x}=\mathbf{1}_{T}\psi^{T}D=\left(D^{T}\otimes\mathbf{1}_{T}\right)\text{vec}\psi^{T}=U\psi$,
where $\psi$ is the vector of steady-states, $\mathbf{1}_{T}$ is
a vector of ones, and $D=(I_{n},\dots,I_{n})$, a $n\times np$ matrix
of $p$ concatenated identity matrices, where $n$ is the number of
time series and $p$ is the number of lags.

\subsection*{Update step for $\Psi$}

Keeping in only the terms in (\ref{eq:VI_all}) containing $\Psi$
(the other terms do not affect the functional form of $q_{\Psi}(\Psi)$)
and using the vectorized form of the likelihood, we obtain

\begin{equation}
\begin{aligned}\begin{aligned}\log q_{\Psi}\left(\Psi\right) & \propto\end{aligned}
 & -\frac{1}{2}\text{E}_{q(\Pi)q(\Sigma)}\left[\left(y_{\Psi}-(I\otimes X_{\Psi})\pi\right)^{T}\left(\Sigma\otimes I\right)^{-1}\left(y_{\Psi}-(I\otimes X_{\Psi})\pi\right)\right]\\
 & -\frac{1}{2}\left(\Psi-\underline{\Psi}\right)^{T}\Omega_{\Psi}^{-1}\left(\Psi-\underline{\Psi}\right)
\end{aligned}
\end{equation}
The first term of this expression can be expanded as

\begin{align}
\begin{array}{cc}
 & -\text{E}_{q(\Pi)q(\Sigma)}\left[y_{\Psi}^{T}\left(\Sigma\otimes I\right)^{-1}y_{\Psi}-y_{\Psi}^{T}\left(\Sigma\otimes I\right)^{-1}(I\otimes X_{\Psi})\pi\right.\\
 & \left.-\pi^{T}(I\otimes X_{\Psi})^{T}(\Sigma\otimes I)y_{\Psi}+\pi^{T}(I\otimes X_{\Psi})^{T}\left(\Sigma\otimes I\right)^{-1}(I\otimes X_{\Psi})\pi\right].
\end{array}\label{eq:PsiDist}
\end{align}
The last term in this expression can be rewritten as

\begin{multline}
E_{q(\Pi)q(\Sigma)}\left[\text{vec}\left(X_{\Psi}\right)^{T}\left(\Pi^{T}\otimes I\right)^{T}\left(\Sigma\otimes I\right)^{-1}\left(\Pi^{T}\otimes I\right)\text{vec}(X_{\Psi})\right]\\
=\text{vec}\left(X_{\Psi}\right)^{T}E_{q(\Pi)q(\Sigma)}\left[\left(\Pi^{T}\otimes I\right)^{T}\left(\Sigma\otimes I\right)^{-1}\left(\Pi^{T}\otimes I\right)\right]\text{vec}(X_{\Psi}),\label{eq:prob-part}
\end{multline}
where
\begin{align*}
E_{q(\Pi),q(\Sigma)}\left[\left(\Pi^{T}\otimes I\right)^{T}\left(\Sigma\otimes I\right)^{-1}\left(\Pi^{T}\otimes I\right)\right] & =E_{q(\Pi),q(\Sigma)}\left[\Pi\Sigma^{-1}\Pi^{T}\otimes I\right]\\
 & =E_{q(\Pi)}\left[\Pi E_{q(\Sigma)}\left[\Sigma^{-1}\right]\Pi^{T}\right]\otimes I.
\end{align*}
We will show below that the optimal $q(\Sigma)$ follows an inverse
Wishart distribution  $\Sigma\sim IW$ so \textbf{$E\left[\Sigma^{-1}\right]$}
is known and will be denoted $S_{\Sigma}$. Define $\tilde{\Pi}\equiv\Pi S_{\Sigma}^{1/2}$,
where $S_{\Sigma}^{1/2}$ is any matrix square root. We then get
\begin{equation}
\begin{array}{cc}
E_{q(\Pi)}\left[\Pi E_{q(\Sigma)}\left[\Sigma^{-1}\right]\Pi^{T}\right]= & E_{q(\tilde{\Pi})}\left[\tilde{\Pi}\tilde{\Pi}^{T}\right],\end{array}
\end{equation}
and since $\text{vec}\tilde{\Pi}=\left(S_{\Sigma}^{1/2}\otimes I\right)\text{vec}\Pi$
we have that
\begin{equation}
\begin{array}{cc}
\text{vec}\tilde{\Pi}\sim & N\left(\left(S_{\Sigma}^{1/2}\otimes I\right)\mu_{\Pi},\left(S_{\Sigma}^{1/2}\otimes I\right)\overline{\Omega}_{\Pi}\left(S_{\Sigma}^{1/2}\otimes I\right)^{T}\right)=N\left(\mu_{\tilde{\Pi}},\overline{\Omega}_{\tilde{\Pi}}\right)\end{array}.
\end{equation}
Now, partition $\tilde{\Pi}$ by columns as $\tilde{\Pi}=\left[\tilde{\Pi}_{1},\dots,\tilde{\Pi}_{n}\right]$,
so that
\begin{equation}
\begin{array}{cc}
E_{q(\tilde{\Pi})}\left[\tilde{\Pi}\tilde{\Pi}^{T}\right]=E_{q(\tilde{\Pi})}\left[\sum_{i=1}^{n}\tilde{\Pi}_{i}\tilde{\Pi}_{i}^{T}\right]= & \sum_{i=1}^{n}E_{q(\tilde{\Pi}_{i})}\left[\tilde{\Pi}_{i}\tilde{\Pi}_{i}^{T}\right].\end{array}
\end{equation}
Since
\begin{equation}
\text{Var}\left(\tilde{\Pi}_{i}\right)=E_{q(\tilde{\Pi}_{i})}\left[\tilde{\Pi}_{i}\tilde{\Pi}_{i}^{T}\right]-E_{q(\tilde{\Pi}_{i})}\left[\tilde{\Pi}_{i}\right]E_{q(\tilde{\Pi}_{i})}\left[\tilde{\Pi}_{i}^{T}\right]
\end{equation}
we have that
\begin{equation}
E_{q(\tilde{\Pi}_{i})}\left[\tilde{\Pi}_{i}\tilde{\Pi}_{i}^{T}\right]=\bar{\Omega}_{ii}+\mu_{i}\mu_{i}^{T}
\end{equation}
using the partition $\mu_{\tilde{\Pi}}=\left(\mu_{1},\dots,\mu_{n}\right)^{T}$,
where $\mu_{i}$ is a $np$-dimensional vector of means of the $i$:th
column of $\tilde{\Pi}$ with the corresponding ($np\times np$) covariance
matrix $\overline{\Omega}_{ii}$. For notational convenience we define
$A_{\Pi\Sigma}\equiv E_{q(\Pi)}\left[\Pi E_{q(\Sigma)}\left[\Sigma^{-1}\right]\Pi^{T}\right]\otimes I$.
Plugging this into (\ref{eq:prob-part}), we obtain
\begin{equation}
\begin{array}{cc}
 & \left(\text{vec}X-\text{vec}\Psi_{x}\right)^{T}A_{\Pi\Sigma}\left(\text{vec}X-\text{vec}\Psi_{x}\right)\propto\psi^{T}U^{T}A_{\Pi\Sigma}U\psi-2x^{T}A_{\Pi\Sigma}U\psi.\end{array}\label{eq:A}
\end{equation}

Looking now at the middle terms of (\ref{eq:PsiDist}) we have
\begin{align*}
E_{q(\Pi),q(\Sigma)}\left[y_{\Psi}^{T}\left(\Sigma\otimes I\right)^{-1}(I\otimes X_{\Psi})\pi\right] & =y_{\Psi}^{T}\left(S_{\Sigma}\otimes I\right)(I\otimes X_{\Psi})\mu_{\Pi}\\
 & =\left(y-vec(\Psi_{y})\right)^{T}\left(S_{\Sigma}\otimes I\right)\left(M_{\Pi}^{T}\otimes I\right)\left(x-vec(\Psi_{x})\right)\\
 & \propto-y^{T}B_{\Pi\Sigma}vec(\Psi_{x})-vec(\Psi_{y})^{T}B_{\Pi\Sigma}x+vec(\Psi_{y})^{T}B_{\Pi\Sigma}vec(\Psi_{x})\\
 & =-y^{T}B_{\Pi\Sigma}U\psi-\psi{}^{T}V^{T}B_{\Pi\Sigma}x+\psi{}^{T}V^{T}B_{\Pi\Sigma}U\psi\\
 & =\psi{}^{T}V^{T}B_{\Pi\Sigma}U\psi-\left(y^{T}B_{\Pi\Sigma}U+x^{T}B_{\Pi\Sigma}^{T}V\right)\psi,
\end{align*}
where $M_{\Pi}$ is a matrix of same dimension as $\Pi$ such that:
$\text{vec}M_{\Pi}=\mu_{\Pi}=\text{E}\left[\text{vec}\Pi\right]$
and we use the short-hand notation $B_{\Pi\Sigma}\equiv\left(S_{\Sigma}\otimes I\right)\left(M_{\Pi}^{T}\otimes I\right)$.
Doing the same thing for the other middle term of (\ref{eq:PsiDist})
and rearranging we get
\[
E_{q(\Pi)q(\Sigma)}\left[\pi^{T}(I\otimes X_{\Psi})^{T}(\Sigma\otimes I)y_{\Psi}\right]\propto\psi^{T}U^{T}B_{\Pi\Sigma}^{T}V\psi-\left(x^{T}B_{\Pi\Sigma}^{T}V+y^{T}B_{\Pi\Sigma}U\right)\psi.
\]
Combining the two middle terms of (\ref{eq:PsiDist}) gives
\begin{equation}
\psi\left(U^{T}B_{\Pi\Sigma}^{T}V+V^{T}B_{\Pi\Sigma}U\right)\psi-2\left(x^{T}B_{\Pi\Sigma}^{T}V+y^{T}B_{\Pi\Sigma}U\right)\psi.\label{eq:B}
\end{equation}

Finally, the last term of (\ref{eq:PsiDist}) is given by:

\begin{equation}
\begin{array}{cc}
E_{q(\Pi)q(\Sigma)}\left[y_{\Psi}^{T}\left(\Sigma\otimes I\right)^{-1}y_{\Psi}\right] & =\left(y-vec(\Psi_{y})\right)^{T}\left(E\left[\Sigma^{-1}\right]\otimes I\right)\left(y-vec(\Psi_{y})\right)\\
 & \propto vec(\Psi_{y})^{T}\left(S_{\Sigma}\otimes I\right)vec(\Psi_{y})-2y^{T}\left(S_{\Sigma}\otimes I\right)vec(\Psi_{y})\\
 & =\psi{}^{T}V^{T}\left(S_{\Sigma}\otimes I\right)V\psi-2y^{T}\left(S_{\Sigma}\otimes I\right)V\psi.
\end{array}\label{eq:C}
\end{equation}
If we put together (\ref{eq:A}), (\ref{eq:B}), (\ref{eq:C}) and
the prior, and match it to a normal distribution we get that $\psi$
is multivariate normal with

\begin{equation}
\begin{array}{cc}
\Sigma_{\psi}= & \left[U^{T}A_{\Pi\Sigma}U+V^{T}\left(S_{\Sigma}\otimes I\right)V-\left(U^{T}B_{\Pi\Sigma}^{T}V+V^{T}B_{\Pi\Sigma}U\right)+\underline{\Omega}_{\psi}^{-1}\right]^{-1}\\
\mu_{\psi}= & \Sigma_{\psi}\left[y^{T}\left(S_{\Sigma}\otimes I\right)V-x^{T}B_{\Pi\Sigma}^{T}V-y^{T}B_{\Pi\Sigma}U+x^{T}A_{\Pi\Sigma}U+\underline{\psi}\underline{\Omega}_{\psi}^{-1}\right].
\end{array}
\end{equation}

\subsection*{Updating equation for $\Pi$}

Similarly as in the update step for $\Psi$ we use the vectorized
notation and we only keep the terms in (\ref{eq:VI_all}) containing
$\Pi$, we then obtain

\[
\begin{aligned}\begin{aligned}\log q_{\Pi}\left(\Pi\right) & \propto\end{aligned}
 & -\frac{1}{2}\text{E}_{q(\Psi)q(\Sigma)}\left[\left(y_{\Psi}-(I\otimes X_{\Psi})\pi\right)^{T}\left(\Sigma\otimes I\right)^{-1}\left(y_{\Psi}-(I\otimes X_{\Psi})\pi\right)\right]\\
 & -\frac{1}{2}\left(\Pi-\underline{\Pi}\right)^{T}\Omega_{\Pi}^{-1}\left(\Pi-\underline{\Pi}\right).
\end{aligned}
\]
Where the first term can be expanded in the same way as in (\ref{eq:PsiDist}),
we get

\begin{equation}
\begin{array}{cc}
 & -\text{E}_{q(\Psi)q(\Sigma)}\left[y_{\Psi}^{T}\left(\Sigma\otimes I\right)^{-1}y_{\Psi}-y_{\Psi}^{T}\left(\Sigma\otimes I\right)^{-1}(I\otimes X_{\Psi})\pi\right.\\
 & \left.-\pi^{T}(I\otimes X_{\Psi})^{T}(\Sigma\otimes I)y_{\Psi}+\pi^{T}(I\otimes X_{\Psi})^{T}\left(\Sigma\otimes I\right)^{-1}(I\otimes X_{\Psi})\pi\right]\\
\propto & -\text{E}_{q(\Psi)q(\Sigma)}\left[\pi^{T}(I\otimes X_{\Psi})^{T}\left(\Sigma\otimes I\right)^{-1}(I\otimes X_{\Psi})\pi-y_{\Psi}^{T}\left(\Sigma\otimes I\right)^{-1}(I\otimes X_{\Psi})\pi\right.\\
 & \left.-\pi^{T}(I\otimes X_{\Psi})^{T}(\Sigma\otimes I)y_{\Psi}\right],
\end{array}\label{eq:PiDist}
\end{equation}
where the first term is independent of $\Pi$ and can be ignored.
The first term in (\ref{eq:PiDist}) can be rewritten as
\begin{equation}
\begin{array}{cc}
= & -\pi^{T}\left(\text{E}_{q(\Sigma)}\left[\Sigma^{-1}\right]\otimes E_{q(\Psi)}\left[\left(X-\Psi\right)^{T}\left(X-\Psi\right)\right]\right)\pi\\
= & -\pi^{T}\left[S_{\Sigma}\otimes\left(X^{T}X+Q_{\Psi_{xx}}-X^{T}M_{\psi}-M_{\psi}^{T}X\right)\right]\pi,
\end{array}\label{eq:Pi1}
\end{equation}
where $Q_{\Psi_{xx}}=E\left[\Psi_{x}^{T}\Psi_{x}\right]=D^{T}E\left[\psi\mathbf{1}_{T}^{T}\mathbf{1}_{T}\psi\right]D=TD^{T}E\left[\psi\psi^{T}\right]D=TD^{T}\left(\Omega_{\psi}+\mu_{\psi}\mu_{\psi}^{T}\right)D$,
where $\Omega_{\psi}$ and $\mu_{\psi}$ are the covariance matrix
and mean vector of the distribution $q(\Psi)$.

Looking at the two last terms of (\ref{eq:PiDist}) we have

\begin{equation}
\begin{aligned}E_{q(\Psi)q(\Sigma)}\left[2y_{\Psi}^{T}\left(\Sigma\otimes I\right)^{-1}(I\otimes X_{\Psi})\pi\right]= & 2E_{q(\Psi)q(\Sigma)}\left[\left(y-\text{vec}\Psi_{y}\right)^{T}\left\{ \left(\Sigma^{-1}\otimes X\right)-\left(\Sigma^{-1}\otimes\Psi_{x}\right)\right\} \right]\pi\\
= & 2\biggl(y^{T}\left(S_{\Sigma}\otimes X\right)-y^{T}\left(S_{\Sigma}\otimes M_{\Psi}\right)\\
 & -\mu_{\psi_{y}}^{T}\left(S_{\Sigma}\otimes X\right)+\text{vec}\left(Q_{\Psi_{xy}}S_{\Sigma}\right)\biggr)\pi,
\end{aligned}
\label{eq:Pi2}
\end{equation}

where $Q_{\Psi_{xy}}$ is defined in a similar way as $Q_{\Psi_{xx}}$.

Combining (\ref{eq:Pi1}) and (\ref{eq:Pi2}) with the prior and matching
it to a normal distribution, we get that $\pi$ is multivariate normal
with
\begin{equation}
\begin{array}{cc}
\Omega_{\pi}^{-1}= & S_{\Sigma}\otimes\left(X^{T}X+Q_{\Psi_{xx}}-X^{T}M_{\psi}-M_{\psi}^{T}X\right)+\underline{\Omega}_{\pi}^{-1}\\
\mu_{\pi}= & \Omega_{\pi}\left(y^{T}\left[\left(S_{\Sigma}\otimes X\right)-\left(S_{\Sigma}\otimes M_{\Psi}\right)\right]-\mu_{\psi}\left(S_{\Sigma}\otimes X\right)+\text{vec}\left(Q_{\Psi_{yx}}S_{\Sigma}\right)+\underline{\pi}^{T}\underline{\Omega}_{\pi}^{-1}\right).
\end{array}
\end{equation}

\subsubsection*{Update for $\Sigma$}

Collecting the terms that with $\Sigma$, we obtain 

\begin{equation}
\begin{aligned}\log q_{\Sigma}\left(\Sigma\right)\propto & -\text{E}_{q(\Psi)q(\Pi)}[T\log|\Sigma|+tr\left\{ \left(Y_{\psi}-X_{\psi}\Pi\right)\Sigma^{-1}\left(Y_{\psi}-X_{\psi}\Pi\right)^{T}\right\} \\
+ & \left(\underline{\nu}+n+1\right)\log|\Sigma|+tr\left(\underline{S}\Sigma^{-1}\right)\\
= & -\left\{ tr\left[\Sigma^{-1}\tilde{S}\right]+\left(T+\underline{\nu}+n+1\right)\log|\Sigma|+tr\left(\underline{S}\Sigma^{-1}\right)\right\} \\
= & -\left\{ \left(T+\underline{\nu}+n+1\right)\log|\Sigma|+tr\left[\left(\underline{S}+\tilde{S}\right)\Sigma^{-1}\right]\right\} .
\end{aligned}
\label{eq:SigDist}
\end{equation}
We can immediately match (\ref{eq:SigDist}) to the inverse Wishart
distribution with $\overline{\nu}=T+\underline{\nu}$ degrees of freedom
and with the scale matrix $\overline{S}=\underline{S}+\tilde{S}$,
i.e. $\Sigma\sim IW\left(\overline{\nu},\overline{S}\right)$. $\tilde{S}$
can be expanded as 

\begin{equation}
\begin{array}{cc}
\tilde{S}= & E_{\Pi,\Psi}\left[\left(Y_{\psi}-X_{\psi}\Pi\right)^{T}\left(Y_{\psi}-X_{\psi}\Pi\right)\right]\\
= & E_{\Pi,\Psi}\biggl[\left(Y-\Psi\right)^{T}\left(Y-\Psi\right)-\left(Y-\Psi\right)^{T}\left(X-\Psi\right)\Pi\\
 & -\Pi^{T}\left(X-\Psi\right)^{T}\left(Y-\Psi\right)+\Pi^{T}\left(X-\Psi\right)^{T}\left(X-\Psi\right)\Pi\biggr].
\end{array}
\end{equation}
The first term in this expression can be rewritten as
\begin{equation}
\begin{array}{cc}
E_{q(\Pi)q(\Psi)}\left[\Pi^{T}\left(X-\Psi\right)^{T}\left(X-\Psi\right)\Pi\right]= & E_{q(\Pi)}\left[\Pi^{T}E_{q(\Psi)}\left[\left(X-\Psi\right)^{T}\left(X-\Psi\right)\right]\Pi\right]\\
= & E_{q(\Pi)q(\Psi)}\left[\Pi^{T}B^{1/2}B^{1/2}\Pi\right]\\
= & E_{q(\overline{\Pi})}\left[\bar{\Pi}^{T}\bar{\Pi}\right]=Q_{\bar{\Pi}},
\end{array}
\end{equation}
where use the same idea as in the update step for $\Psi$, where $B^{1/2}$
denotes any matrix square root of $E_{q(\Psi)}\left[\left(X-\Psi\right)^{T}\left(X-\Psi\right)\right]$.
Furthermore,
\begin{equation}
\begin{array}{cc}
B= & E\left[\left(X-\Psi\right)^{T}\left(X-\Psi\right)\right]\\
= & X^{T}X-X^{T}M_{\Psi_{x}}-M_{\Psi_{x}}^{T}X+Q_{\Psi_{xx}},
\end{array}
\end{equation}

where $M_{\Psi_{x}}$ and $Q_{\Psi_{xx}}$ has already been calculated
in the update step for $\Pi$. Note that $Q_{\bar{\Pi}}$ can be calculated
in the same way as $Q_{\tilde{\Pi}}$ in the update step for $\Psi$,
but since the transpose is reversed we will instead have a partition
by rows, i.e. $\mu_{\bar{\Pi}}=\left(\mu_{1},\dots,\mu_{np}\right)^{T}$,
where $\mu_{i}$ is a $n$-dimensional vector of means with the corresponding
($n\times n$) covariance matrix $\overline{\Omega}_{ii}$. Note that
when calculating the transformed covariance matrix it is important
to reorder the elements such that we get the covariance matrix of
$\text{vec}\Pi^{T}$ rather than the covariance matrix of $\text{vec}\Pi$.

For the middle terms we have
\begin{equation}
\begin{array}{cc}
E\left[\left(Y-\Psi\right)^{T}\left(X-\Psi\right)\Pi\right]= & \left(Y^{T}X-E\left[\Psi_{y}^{T}\right]X-Y^{T}E\left[\Psi_{x}\right]+E\left[\Psi_{y}^{T}\Psi_{x}\right]\right)E\left[\Pi\right]\\
= & \left(Y^{T}X-M_{\Psi_{y}}^{T}X-Y^{T}M_{\Psi_{x}}+Q_{\Psi_{yx}}\right)M_{\Pi}=C_{2}M_{\Pi}.
\end{array}
\end{equation}
And the first term is given by
\begin{equation}
\begin{array}{cc}
E\left[\left(Y-\Psi\right)^{T}\left(Y-\Psi\right)\right]= & Y^{T}Y-Y^{T}E\left[\Psi_{y}\right]-E\left[\Psi_{y}^{T}\right]Y+E\left[\Psi_{y}^{T}\Psi_{y}\right]\\
= & Y^{T}Y-Y^{T}M_{\Psi_{y}}-M_{\Psi_{y}}^{T}Y+Q_{\Psi_{yy}}=C_{1}.
\end{array}
\end{equation}

Combining the terms gives us
\begin{equation}
\begin{array}{cc}
\tilde{S}= & C_{1}-C_{2}M_{\Pi}-M_{\Pi}^{T}C_{2}^{T}+Q_{\bar{\Pi}}\end{array}.
\end{equation}

\section{Additional results for the US macroeconomic application\label{sec:List-of-figures}}

\begin{figure}[H]
\centering{}\includegraphics[scale=0.5]{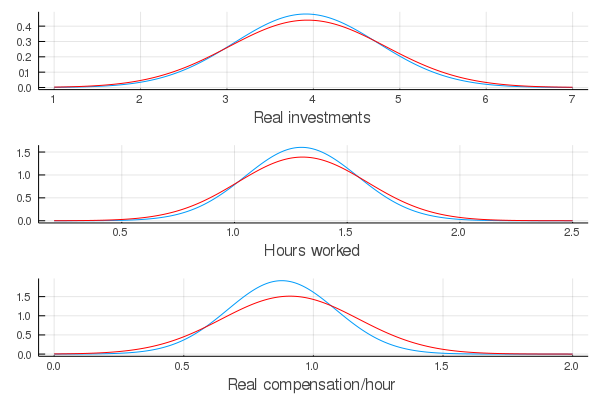}\caption{Posterior distribution for $\Psi$, medium sized model.\label{fig:real_psi-1}}
\end{figure}

\begin{figure}[H]
\centering{}\includegraphics[scale=0.5]{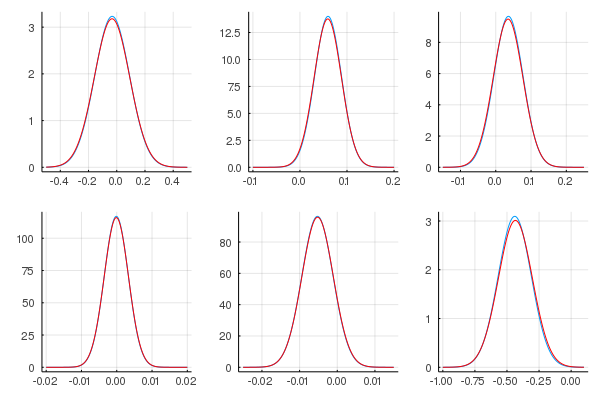}\caption{Posterior distributions for some randomly selected $\Pi$-coefficients
for the medium sized model.\label{fig:real_Pi}}
\end{figure}

\begin{figure}[H]
\centering{}\includegraphics[scale=0.55]{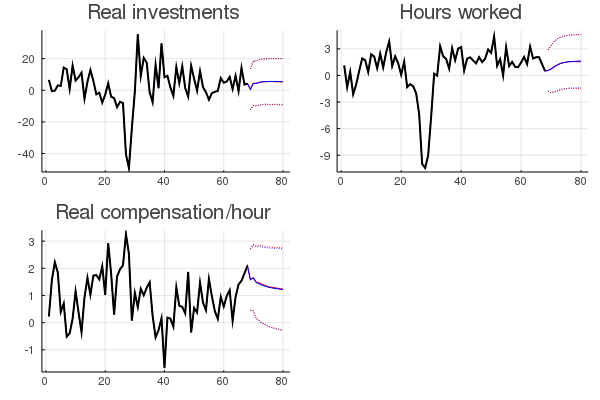}\caption{Out of sample forecasts for the medium sized model.\label{fig:real_forecast_append}}
\end{figure}

\begin{figure}[H]
\centering{}\includegraphics[scale=0.55]{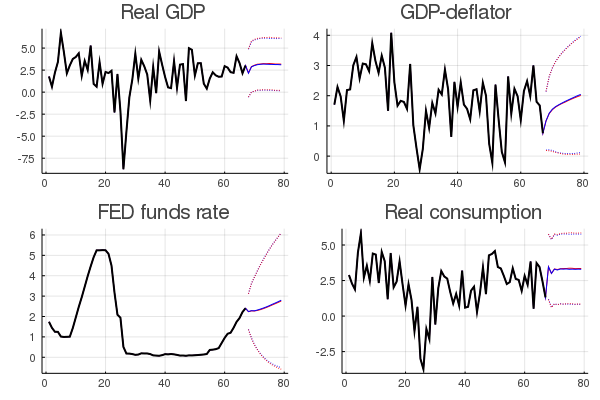}\caption{Out of sample forecasts using the large BVAR model.\label{fig:Large_forecasts}}
\end{figure}

\end{document}